\documentclass[10pt]{iopart}
\usepackage{times}
\usepackage{graphicx}
\usepackage{subfigure}

\begin{document}
\title[Inferring the effective thickness of polyelectrolytes from
stretching measurements]{Inferring the effective thickness of polyelectrolytes from
stretching measurements at various ionic strengths: applications to
DNA and RNA}

\author{Ngo Minh Toan$^{1,2}$ and Cristian Micheletti$^1$}
\eads{\mailto{ngo@sissa.it}, \mailto{michelet@sissa.it}}
\address{$^1$ International School for Advanced Studies (S.I.S.S.A.)
and INFM, Via Beirut 2-4, 34014 Trieste, Italy}
\address{$^2$ Institute of Physics, 10 Dao Tan, Hanoi, Vietnam}

\date{\today}

\begin{abstract}{By resorting to the thick-chain model we discuss how
the stretching response of a polymer is influenced by the
self-avoidance entailed by its finite thickness. The characterization
of the force versus extension curve for a thick chain is carried out
through extensive stochastic simulations. The computational results
are captured by an analytic expression that is used to fit
experimental stretching measurements carried out on DNA and
single-stranded RNA (poly-U) in various solutions.  This strategy
allows us to infer the apparent diameter of two biologically-relevant
polyelectrolytes, namely DNA and poly-U, for different ionic
strengths. Due to the very different degree of flexibility of the two
molecules, the results provide insight into how the apparent diameter
is influenced by the interplay between the (solution-dependent) Debye
screening length and the polymers' ``bare'' thickness. For DNA, the
electrostatic contribution to the effective radius, $\Delta$, is found
to be about 5 times larger than the Debye screening length,
consistently with previous theoretical predictions for highly-charged
stiff rods. For the more flexible poly-U chains the electrostatic
contribution to $\Delta$ is found to be significantly smaller than the
Debye screening length.}
\end{abstract}

\maketitle

\section*{Introduction}

In recent years, the remarkable advancement of single-molecule
manipulation techniques has made possible to characterise with great
accuracy how various biopolymers respond to mechanical stretching
\cite{bustamante03,busta,single2,strick,block1,williams,linke,
fernan,gaub99,marsza2,clarke1,clarke2,wang,strick2000}. The wealth of
collected experimental data have constituted and still represent an
invaluable and challenging benchmark for models of polymers'
elasticity \cite{Boal,pincus,WLC2,cieplak10,schulten1,bensimon98,
TCstretch05,Hamed,netz2,pianaMDstretch}. The interpretation of
single-molecule stretching experiments often relies on one-dimensional
non-self-avoiding models of polymers. It is physically appealing that
the schematic nature of such descriptions often conjugates with the
capability of reproducing well experimental measurements. Two notable
instances are represented by the freely-jointed chain
\cite{flory,Boal} and the worm-like chain models which, in their
original or extended forms, constitute the most commonly-used
theoretical frameworks for biopolymers'
stretching\cite{WLC1,WLC2,podg01}. Both models are endowed with a
parameter, the Kuhn length or the persistence length, that provides a
phenomenological measure of the polymer stiffness and that is obtained
by fitting the experimental data. It is important to notice, however,
that it is possible to go beyond this phenomenological approach
and connect the persistence length to the fundamental structural
properties of a polymer. A strong indication of the feasibility of
this scheme is provided by the fact that, for a large number of
biopolymers, the observed persistence length shows an approximate
quartic dependence on the polymer diameter, as predicted for ``ideal''
elastic rods\cite{Boal}.

From this perspective it appears natural to investigate in detail the
connection between structural properties and stretching response of
biopolymers. We have recently pursued this objective by modelling
explicitly the intrinsic thickness of a homo-polymer (treated as a
tube with uniform cross-section) and characterising the stretching
response \cite{TCstretch05}. The theoretical and numerical results
were employed in an appealing chemico-physical framework where the
diameter of a biopolymer was not probed directly but inferred through
the mere knowledge of the stretching response. The adopted thick-chain
model \cite{Thickness1,opthelix,Thickness3}, briefly outlined in the
next section, was used to fit stretching measurements obtained for a
variety of biopolymers: DNA \cite{block1}, the PEVK-domain of the
titin protein \cite{linke,fernan} and cellulose
\cite{marsza1,marsza2}. For uncharged polymers, such as titin and
cellulose, the effective diameter recovered from fitting the
force-extension curves were very consistent with the stereochemical
ones, thereby validating the thick-chain model approach
\cite{TCstretch05}. Even more interesting is the case of polymers
possessing a substantial linear charge density, such as DNA and RNA
which will be the focus of the present study. The properties of
polyelectrolytes, in fact, depend very strongly on the electrostatic
screening provided by the ions present in solution. The influence of
the electrostatic screening on the behaviour of polyelectrolytes has
been extensively investigated both experimentally and theoretically
\cite{Odijk77,SF77,stigter77,Tricot84,Reed91,Ullner97,thiru}. From the
latter perspective, it is customary to introduce apparent (or
effective) physico-chemical parameters to describe the properties of a
polyelectrolyte in a given solution with reference to the uncharged
polymer case. For example, in the context of elasticity, one
introduces an effective (solution-dependent) bending rigidity to
account for the additional electrostatic contribution to the ``bare''
persistence length of the hypothetically-neutral polymer
\cite{Ullner97,thiru}. Also, in the context of colloidal dispersions
of stiff polyelectrolytes, one can describe the polymer as uncharged
cylinders and resort to the theory for second virial coefficient to
derive its solution-dependent effective
diameter\cite{stigter77,odijk90,stigter93}. For DNA in solutions of low ionic
strength, both the effective persistence length and effective diameter
can exceed by several factors the bare ones. So far, these effective
DNA properties have been typically probed by distinct
methodologies. For example stretching experiments were employed to
establish the dependence of the the persistence length on ionic
strength \cite{WLC4,williams} while measurements of second virial
coefficients, knotting probabilities or braiding properties were used
for the effective
diameter\cite{Nicolai89,stigter93,Sottas99,Cozzarelli,bensimon_braid}).

The thick-chain framework is used to obtain, starting from stretching
measurements data, the effective diameter of a polyelectrolyte and to
further relate it to its the effective persistence length. Besides the
implications in the general context of polyelectrolytes the proposed
method can be used to establish the effective structural parameters to
be used in coarse-grained studies of looping, knotting and packaging
of biopolymers~\cite{harvey,dnapack}.

\section*{The thick chain model}

To characterise the stretching response of a polymer with finite
thickness we shall view the latter as a tube with a uniform circular
section in the plane perpendicular to the tube centreline. The chain
thickness, $\Delta$, is defined as the radius of the circular
section. Several frameworks have been introduced to capture the
uniform thickness constraint in a way apt for numerical
implementation. These approaches typically rely on a discretised
representation of the thick chain
\cite{buck1,Thickness2,rawd99,Thickness1,raw2000}. In this study we
shall employ the piece-wise linear modelling of the chain centerline
introduced by Gonzalez and Maddocks \cite{Thickness1}.

We shall indicate with $\Gamma$ the centerline of the chain consisting
of a succession of points $\{ \vec{r}_0, \vec{r}_1,...\}$ equispaced
at distance $a$. We shall further denote with $\vec{b}_i$ the virtual
bond joining the $i$th and $i+1$th points, $\vec{b}_i = \vec{r}_{i+1}
- \vec{r}_i$. In order for the succession of points $\{\vec{r}_i\}$ to
be a viable centerline for a chain of thickness $\Delta$, it is
necessary that the radii $r_{ijk}$ of the circles going through any
triples of points $i$, $j$ and $k$, are not smaller than
$\Delta$. Accordingly, the Hamiltonian for the thick chain (tube)
model can be written as

\begin{equation}
 {\cal H}_{TC}(\Gamma) = \sum_{ijk} V_3(r_{ijk})
\label{eqn:ham}
\end{equation}

\noindent where $V_3$ is the three-body potential used to
enforce the thickness $\Delta$ of the chain
\cite{Thickness1,Thickness2,jstatphys,opthelix,Thickness3,secstr}.  As
anticipated, the argument of $V_3$ is the radius of the circle going
through the triplet of distinct points $i,\ j,\ k$ and has the form
\begin{equation}
V_3(r) = \left\{ 
\begin{array}{l l}
0 & \mbox{if $r > \Delta$,} \\
+\infty & \mbox{otherwise.} 
\end{array}
\right .
\label{eqn:v3}
\end{equation}

\begin{figure}[h]
\begin{center}
\includegraphics[width=2.5in]{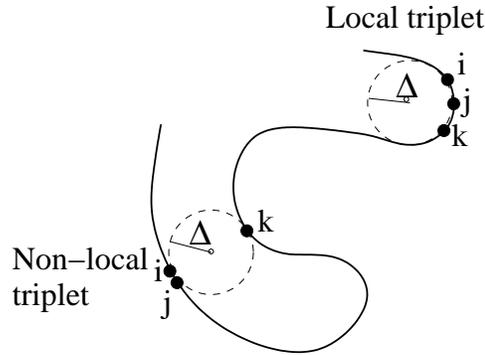}
\end{center}
\caption{The finite thickness introduces steric constraints of local
and non-local character that forbid configurations where the chain
self-intersects. These constraints are conveniently treated
within the three-body prescription of the thick-chain model. Within
this approach the centerline of a viable configuration is such that
the radii, $R_{ijk}$ of the circles going through any triplet of
points on the curve $i,\ j,\ k$ are not smaller than
$\Delta$.}\label{fig:tube}
\end{figure}

Physically, the model of eqn.~\ref{eqn:ham} introduces
conformational restrictions for the centerline that are both local and
non-local in character, as depicted in Fig. \ref{fig:tube}. The local
constraints are those where the triplet $i$, $j$ and $k$ identifies
three consecutive points. The limitation on the radius of the
associated circumcircle reflects the fact that, to avoid
singularities, the local radius of curvature must not be smaller than
$\Delta$. This reflects on the following bound on the angle formed by
two consecutive bonds:

\begin{equation}
{\vec{b}_i \cdot \vec{b}_{i+1} \over a^2} \ge 1 - { a^2 \over
  2\Delta^2}
\label{eqn:angle}
\end{equation}

On the other hand there is also a non-local effect due to the fact
that any two portions of the centerline at a finite arclength
separation cannot interpenetrate. It has been shown in
ref. \cite{Thickness1} that this second effect can be accounted for by
requiring that the minimum radius among circles going through any
triplet of {\em non-consecutive} points, is also greater than (or
equal to) $\Delta$. The seamless way in which the local and non-local
steric effects are accounted for make the model particularly
appealing. Other discrete models relying on pairwise interactions for
the excluded volume (such as the cylindrical model of
refs. \cite{Cozzarelli,bensimon_braid}) may be adopted, though {\em ad
hoc} prescriptions for dealing with e.g. overlapping consecutive units
need to be introduced \cite{RMP}.

In the present context we will consider the application of a
stretching force, $\vec{f}$, to the ends of a chain $\Gamma
={\vec{r_0}, ... , \vec{r}_N}$ of thickness $\Delta$ (the contour
length therefore being $L_c = Na $). The Hamiltonian of
eqn.~\ref{eqn:ham} needs to be complemented with the stretching energy

\begin{equation}
{\cal H} = \sum_{ijk} V_3(r_{ijk})  - \vec{f}\cdot (\vec{r}_N- \vec{r}_0)\ .
\label{eqn:hamstr}
\end{equation}

As customary we shall characterize the force dependence of
the average normalised projection of the end-to-end distance,
$\vec{r}_N- \vec{r}_0$, along the direction of applied force:

\begin{equation}
x = \langle {(\vec{r}_N- \vec{r}_0) \cdot \vec{f} \over N\, a\, |\vec{f} |} \rangle
\label{eqn:stretch}
\end{equation}

\noindent where the brackets denote the canonical ensemble
average. Owing to its self-avoiding nature, the stretching response of
the chain cannot be characterised exactly by available analytical
methods. We shall therefore resort to extensive Monte Carlo samplings,
based on the Metropolis scheme, to evaluate the ensemble averages of
eqn.~\ref{eqn:stretch}.

Besides the numerical study of the tube model subject to
stretching it is interesting to illustrate a simplification of the
model of eqn.~\ref{eqn:ham}, which is amenable to an extensive
analytical characterization. To do so we retain only the {\em local}
thickness constraint and thus end up with a model that is essentially
non-self-avoiding. The simplified nature of this problem, however,
makes it very tractable also in the presence of a bending rigidity
penalty, $\kappa_b$. We shall therefore consider the Hamiltonian

\begin{equation}
{\cal H} = \sum_i V_3(r_{i,i+1,i+2}) - \vec{f} \cdot (\vec{r}_N-
\vec{r}_0) - \kappa_b \sum_i {\vec{b}_i \cdot \vec{b}_{i+1}\over a^2}\ ,
\label{eqn:LTCBR}
\end{equation}

\noindent again we stress that the three-body potential is restricted
only to consecutive (local) triplets. In this form the stretching
response of the model can be characterized exactly both at very low
and very high forces using standard statistical-mechanical
procedures~\cite{angelo,angelothesis}. These two limiting regimes can
be joined together to yield the following approximate expression for
the stretching response of a locally-thick chain with bending rigidity
(LTC+BR, for brevity):

\begin{eqnarray}\label{eq:TC_KP_simplest}
\beta a F &=& {2K}\left( \sqrt{1+\left({1\over
    2K}\right)^2{1\over\left(1-x\right)^2}}-\sqrt{1+\left({1\over
    2K}\right)^2}\right)+ \nonumber \\
&&\left(3{1-y\left(K,\Delta/a\right)\over
            1+y\left(K,\Delta/a\right)}-{1\over
            2K\sqrt{1+\left(1/2K\right)^2}} \right)x
\end{eqnarray}

\noindent where $K = \beta\kappa_b$, $\beta= 1/{K_B\, T}$ is the
inverse Boltzmann factor and
\begin{equation}
y(K,\Delta/a) = \left\{
\begin{array}{l l}
1-{a^2\over 2\Delta^2}\left({1 \over 1- e^{z}} + {1 \over z} \right) & {\Delta\over a} > 0.5\\
\coth(K)-{1\over K} & {\Delta\over a} \le 0.5
\end{array}
\right .
\end{equation}
with $z = {a^2 \over 2\Delta^2}K$. The two cases in the above
equation, reflect the fact that, for $\Delta < a/2$ no restriction
applies to the angle formed by two consecutive virtual bonds since
eqn. (\ref{eqn:angle}) is always satisfied.

Expression \ref{eq:TC_KP_simplest} possesses some noteworthy
limits. First, in the absence of thickness and in the continuum limit
($a \to 0$, $K \to \xi_p/a$, $\xi_p$ being the persistence
length)~\cite{angelo,angelothesis}, the model reduces to the
well-known Marko and Siggia result for the WLC:

\begin{equation}
 f(x) = \frac{k_BT}{\xi_p}\left[\frac{1}{4(1-x)^2}-\frac{1}{4}+
{x}\right] \label{eqn:wlc}
\end{equation}

Secondly, in the absence of both thickness and bending
rigidity one recovers the low- and high-force response of the
freely-jointed chain with Kuhn length equal to $a$:

\begin{equation}
 f(x) \approx \left\{ 
\begin{array}{l l}
	{3k_BT\over a} x, &x\rightarrow 0;\\
	{1\over 1-x} , &x\rightarrow 1.
\end{array}
\right .
\label{eqn:fjc}
\end{equation}

It is of interest also the case of finite thickness $\Delta /a > 0.5$
but no bending rigidity.  In this case one obtains the following
expression for the persistence length:

\begin{equation}
\xi_p = -{a\over \ln{\left(1-{a^2\over 4 \Delta^2}\right)}}\ .
\label{eqn:xip}
\end{equation}

Though this expression does not include the non-local
self-avoiding effects it will be shown later to provide a good
approximation of the persistence length obtained numerically for the
full model of eqn.~\ref{eqn:ham}.

\begin{figure}[h]
 \subfigure[]
 {
   \label{fig:perlengthMC}
   \includegraphics[width=2.5in]{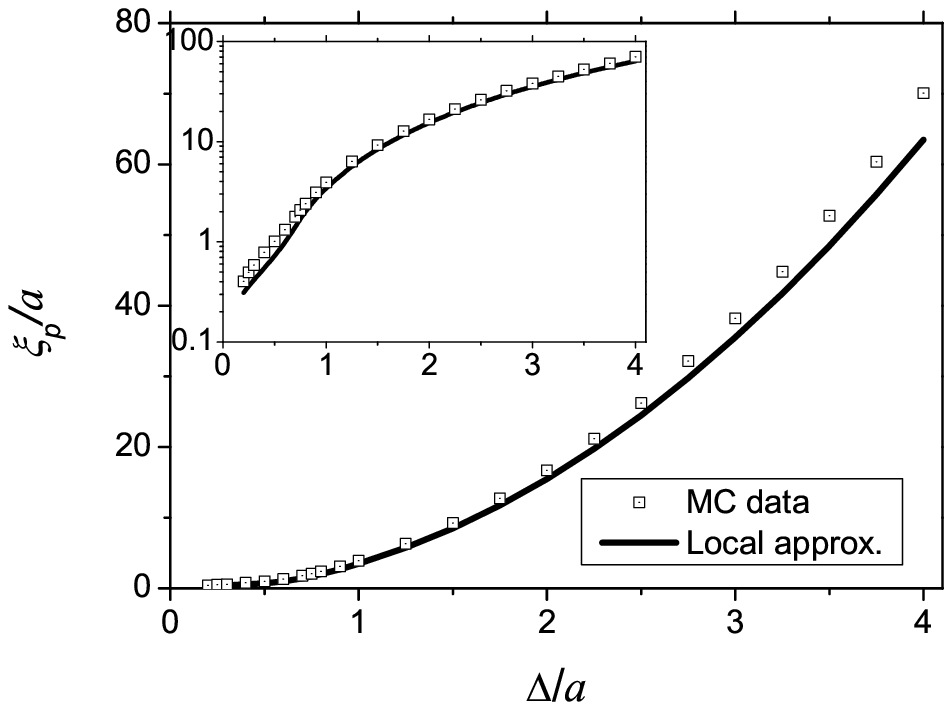}
 }
 \subfigure[]
 {
  \label{fig:ttcorr}
  \includegraphics[width=2.5in]{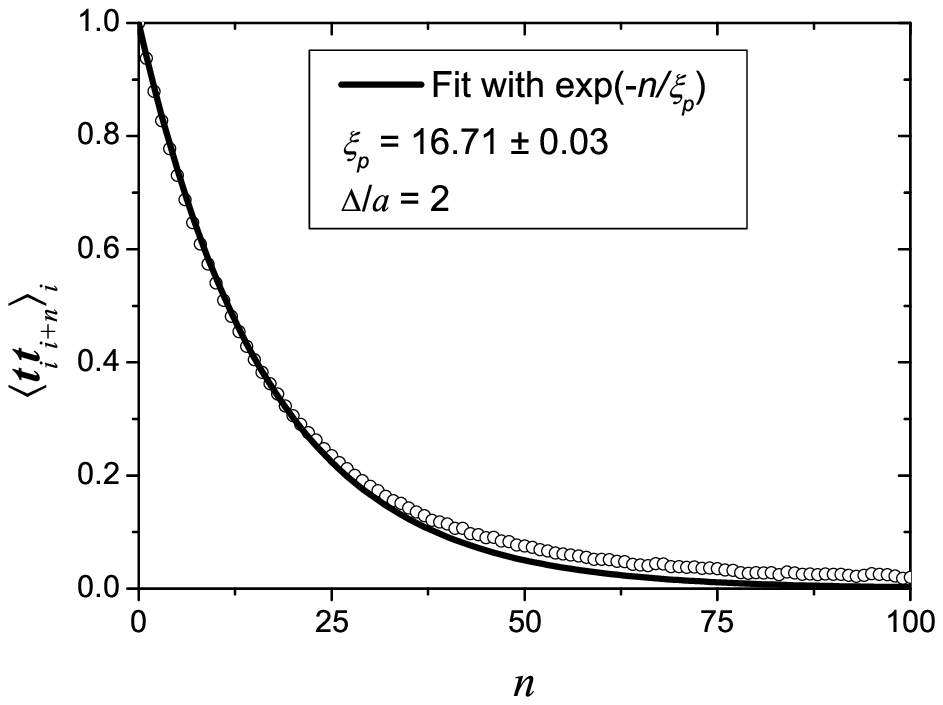}
 }
\caption{(a) Comparison between the persistence length of a thick
chain obtained by Monte Carlo sampling (dotted-square) and that
obtained from the local thickness approximation (solid curve) of
eqn.~(\ref{eqn:xip}). The inset illustrates the limitations of the
local approximation for low values of $\Delta/a$.  (b) Tangent-tangent
correlation (dotted-square) obtained from the analysis of 10$^4$
(uncorrelated) Monte-Carlo-generated chains of 1000 segments and with
$\Delta/a = 2$. The error bars which show the uncertainty of the
plotted values are smaller than the square symbols. The solid curve is
a single exponential fit to the numerical data yielding $\xi_p = 16.71
\pm 0.03$.}
\label{fig:xip-ttcorr}
\end{figure}

We conclude this section by mentioning that for the models of
eqn.~\ref{eqn:ham} and \ref{eqn:LTCBR} the spacing of consecutive
points is constant along the centerline, so that the contour length is
unaffected by the application of forces of arbitrary strength. The
inextensibility property is obviously a simplification of the
behaviour found in naturally-occurring polymers which, at very high
forces can undergo isomerization or structural transitions resulting
in an ``overstretching'' beyond their nominal contour length.  Several
approximate treatments have been developed to correct the stretching
response of inextensible models so to account for overstretching
\cite{odijk1,block1} by adopting additional parameters in the theory.
In the present study we shall keep the number of model parameters to a
minimum and hence postpone to future work the the investigation of the
most suitable way to include overstretching in the TC model.

\section*{Numerical results}

The characterization of the stretching response of the thick chain of
eqn.~\ref{eqn:hamstr} was carried out using a Monte Carlo scheme:
starting from an arbitrary initial chain configuration satisfying the
thickness constraints, the exploration of the available structure
space was done by distorting conformations by means of pivot and
crankshaft moves. Newly-generated structures are accepted/rejected
according to the standard metropolis criterion (the infinite strength
of the three-body penalties of eqn.~\ref{eqn:v3} was enforced by
always rejecting configurations violating the circumradii
constraints).

 The discretization length, $a$, was taken as the unit length in the
problem and several values of $\Delta/a$ were considered, ranging from
the minimum allowed value of 0.5 to the value of 4.0. This upper limit
appears adequate in the present context since the largest nominal
ratio for $\Delta/a$ among the biopolymers considered here is achieved
for dsDNA for which one has $\Delta/a \approx 3.7$ \cite{dnapack}. For
each explored value of $\Delta/a$ considered, we considered chains of
length at least ten times bigger than the persistence length estimated
through eqn. (\ref{eqn:xip}).  The relative elongation of the
chain, was calculated for increasing values of the applied stretching
force (typically about 100 distinct force values were considered).
For each run, after equilibration, we measured the autocorrelation
time and sampled a sufficient number of independent conformations to
achieve a relative error of, at most, $10^{-3}$ on the average chain
elongation. For moderate or high forces this typically entailed the
collection of 10$^4$ independent structures while a tenfold increase
of sampling was required at small forces due to the broad distribution
of the end-to-end separation along the force direction.

We first discuss the results for the persistence length obtained from
the decay of the tangent-tangent correlations measured at zero force
over an ensemble of sampled structures picked at times greater than
the system autocorrelation time. The resulting data are shown in Fig.
\ref{fig:perlengthMC}, along with the curve corresponding to the
approximate expression of eqn. (\ref{eqn:xip}). It can be seen that
the local approximation for the persistence length is very good in the
range $1.0 \le \Delta \le 4 $, where the relative difference from the
value found numerically is typically inferior to 10\%. Significant
relative discrepancies are, instead found as $\Delta$ approaches the
limiting value of 0.5 (though it should be noted, the single
exponential fit suffers from the very rapid decay of the tangent
autocorrelation). In this case, only a narrow range of
values for the angle formed by two consecutive bonds is
forbidden. Consequently, the persistence length is very much affected
by the (non-local) self-avoidance condition that is unaccounted for by
the simple expression of eqn.  (\ref{eqn:xip}). As intuitively
expected the value of $\xi_p$ found numerically is larger than the one
based on the local-thickness approximation.

\begin{figure}[h]
  \subfigure[]
  {
    \label{fig:fvsxMClinear}
    \includegraphics[width=2.5in]{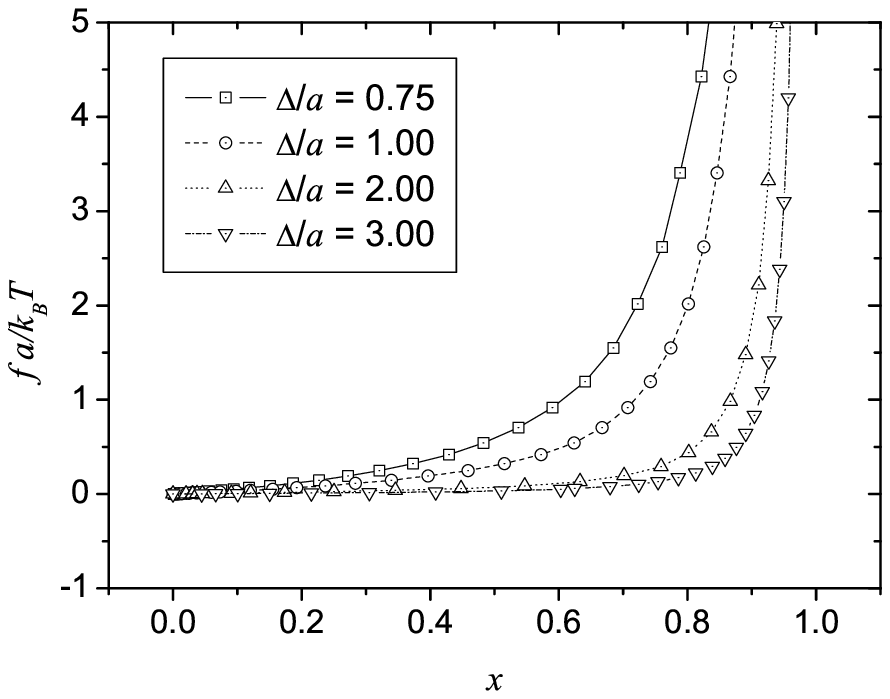}
  }
  \subfigure[]
  {
    \label{fig:fvsxMClog}
    \includegraphics[width=2.5in]{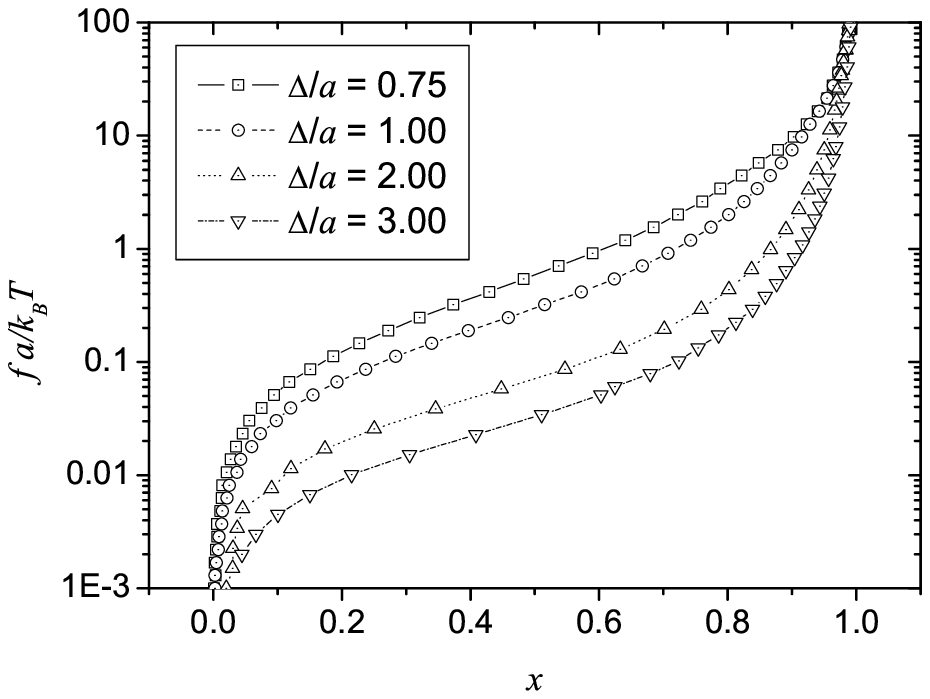}
  }
\caption{ Linear (a) and semi-log (b) plots of the force versus
relative extension curves obtained from Monte Carlo sampling of chains
of relative thickness $\Delta/a =$ 0.75, 1, 2 and 3.}
\label{fig:fvsx}
\end{figure}

\begin{figure}[h]
  \subfigure[]
  {
    \label{fig:FJCWLCregs}
    \includegraphics[width=2.5in]{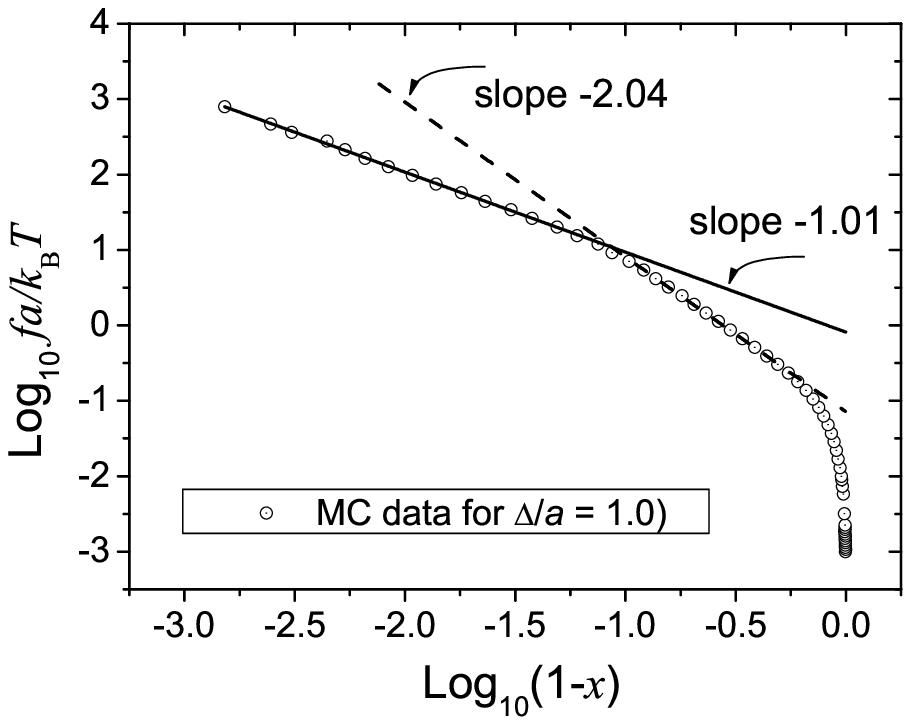}
  }
  \subfigure[]
  {
    \label{fig:PincusRegime}
    \includegraphics[width=2.5in]{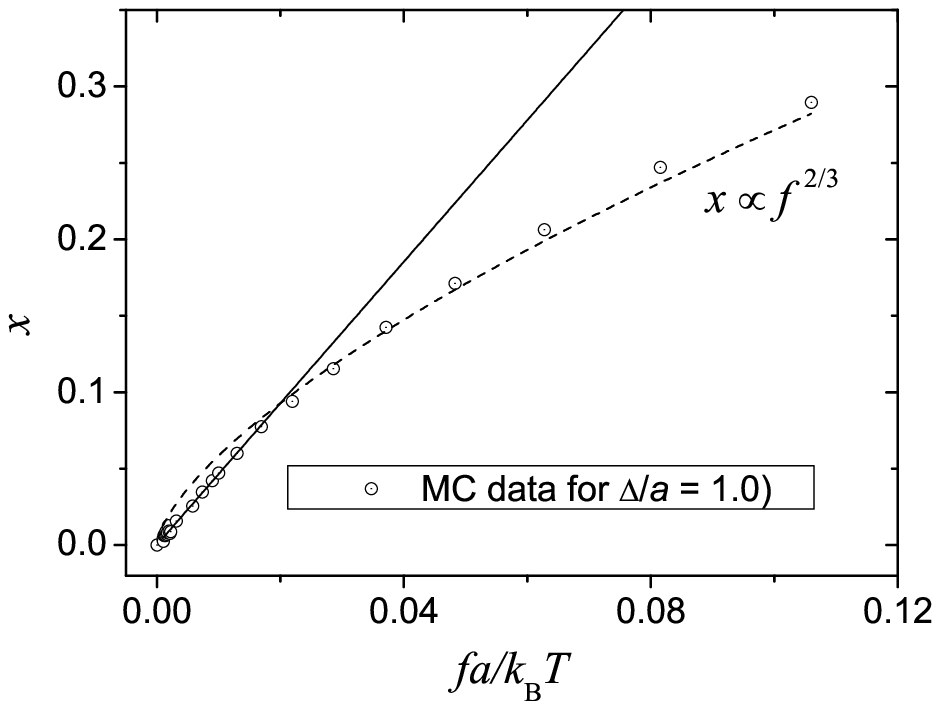}
  }
\caption{(a) Extension versus reduced force for a chain of relative
thickness $\Delta/a=1$. Data points are presented in the
$\left(\log(1-x),\log(f)\right)$ plane to highlight the WLC- and
FJC-like regimes found at moderate and high forces. (b) Illustration
of the low-force crossover from the Hookean regime, $x \propto f$, to
the Pincus one, $x \propto f^{2/3}$ for a chain of 1200 segments and
$\Delta/a=1$.}\label{fig:TCregimes}
\end{figure}

The analysis of the numerical results revealed several different
stretching regimes in the elastic response of a thick chain. These are
best discussed in connection with analogous regimes found in the two
standard reference models, the freely-jointed chain (FJC) and the
worm-like chain (WLC). We first point out that for both these models,
as well as the generalisation of eqn.~\ref{eq:TC_KP_simplest} the
relative elongation, $x$, depends linearly on the applied force,
$f$. This result holds also for the thick chain model. However, due to
the inclusion of the self-avoidance effect in the TC (absent in both
the FJC and WLC), the Hookean relationship between $f$ and $x$
disappears in the limit of long polymer chains in favour of the Pincus
regime, $f\sim x^{1/(1/\nu-1)}$, $\nu \sim 3/5$ being the critical
exponent for self-avoiding polymers in three dimensions
\cite{flory,pincus,lam}, see Fig. \ref{fig:TCregimes}.

For intermediate forces the Pincus behaviour is found to be followed
by a second regime characterised by the same scaling relation found in
the WLC at high forces, $f\sim (1-x)^{-2}$. As shown in Fig.
\ref{fig:TCregimes}, at still higher forces the same scaling relation
found in the FJC is observed, $f\sim (1-x)^{-1}$.  Physically, the
first two regimes are determined by self-avoidance and chain stiffness
or persistence length while the last regime is ascribable to the
discrete nature of the chain \cite{netz,nelson,angelo,lamura}.

In order to apply the thick-chain model to contexts where
experimental data are available we have analysed the data of the
numerical simulations with the purpose of extracting an analytical
expression capturing the observed functional dependence of $f$ on
$\Delta$ and $a$. For any value of $a$ and $\Delta$ the sought
expression should reproduce the succession of the three regimes
discussed above. Among several trial formulae we found, {\em a
posteriori}, that the best interpolation was provided by the following
expression,

\begin{equation}
f(x)=\frac{k_BT}{a (1 -x)}
\tanh{\left(\frac{k_1x^{3/2}+k_2x^2+k_3x^3}{1-x}\right)}\ ,
\label{analytic_approximation}
\end{equation}

\noindent where the parametric dependence on $\Delta$ and $a$ is
carried by the following expressions for the $k$'s,

\begin{eqnarray}
k_1^{-1} & = & -0.28394 + 0.76441\, {\Delta}/{a} + 0.31858\, {\Delta^2}/{a^2},\\
k_2^{-1} & = & +0.15989 - 0.50503\, {\Delta}/{a} - 0.20636\, {\Delta^2}/{a^2},\\
k_3^{-1} & = & -0.34984 + 1.23330\, {\Delta}/{a} + 0.58697\,
{\Delta^2}/{a^2}.
\end{eqnarray}

\begin{figure}[h]
  \subfigure[]
  {
    \label{fig:refit075}
    \includegraphics[width=2.5in]{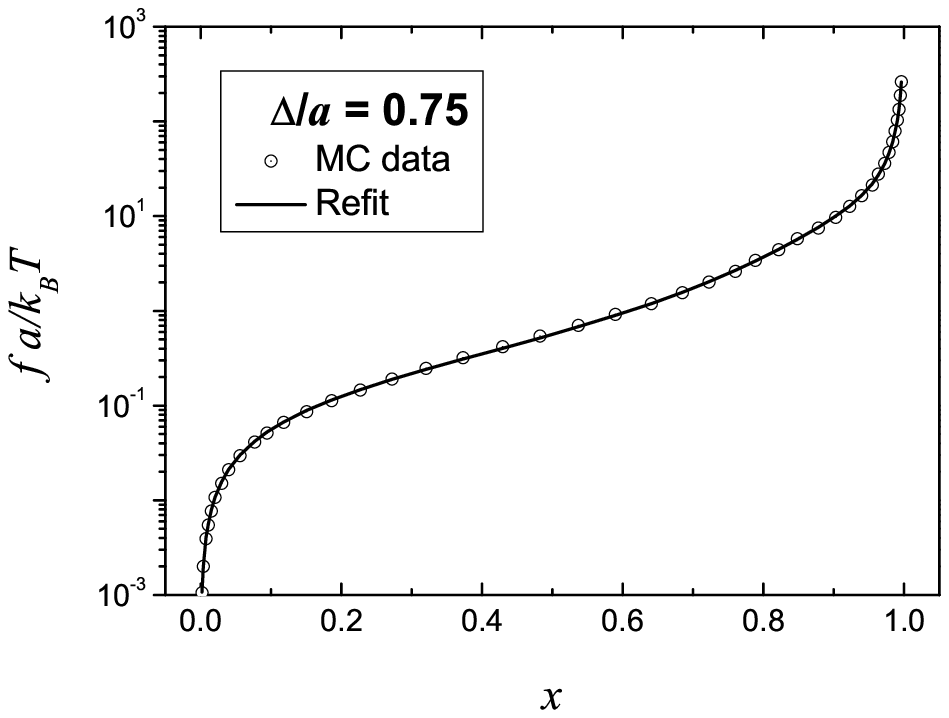}
  }
  \subfigure[]
  {
    \label{fig:refit400}
    \includegraphics[width=2.5in]{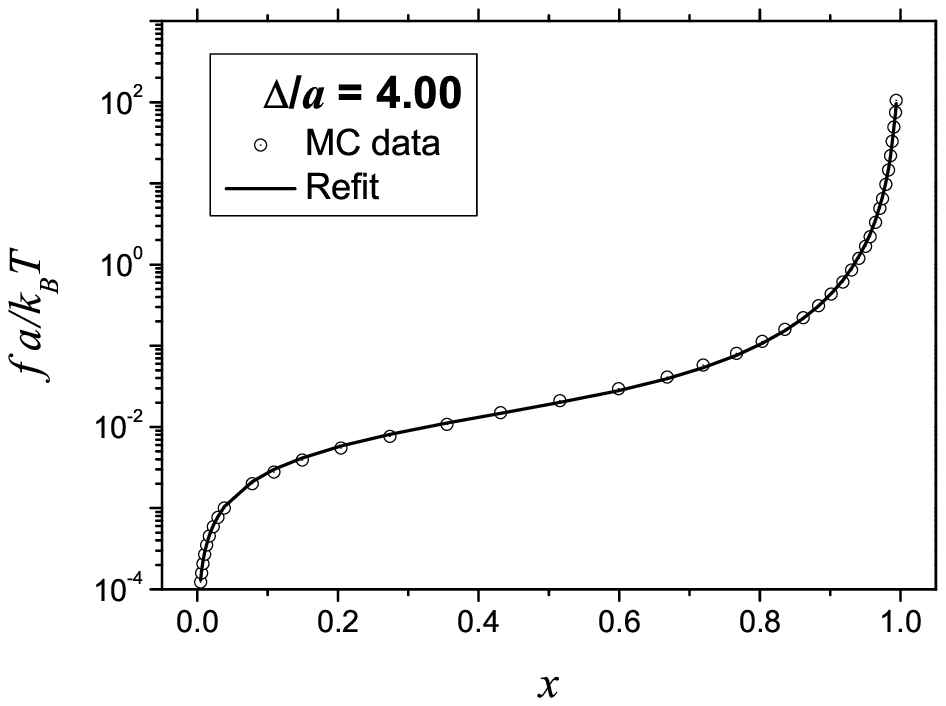}
  }
\caption{Illustration of the performance of the parametric
expression of eqn.~(\ref{analytic_approximation}) in reproducing the
stretching response obtained numerically for chains of relative
thickness $\Delta/a = 0.75$ and $\Delta/a = 4.0$ (figures (a) and
(b), respectively). In both cases the discrepancy between the
computed and parametrised values is about 1\% on average (and always
smaller than 5\%).} \label{fig:refits}
\end{figure}

Within the explored ranges of $\Delta$ and $a$, the relative extension
obtained from eq. (\ref{analytic_approximation}) differs on average by
1\% (and at most by 5\%) from the true values at any value of the
applied force, as shown in Fig. \ref{fig:refits}.

\section*{Applications to experimental data and discussions}

We shall now discuss the application of the TC model to data sets
obtained in DNA and RNA stretching experiments carried out for various
[Na$^+$] concentrations. In particular, the data for dsDNA are taken
from ref.~\cite{busta} (solutions of 10, 1.0 and 0.1 mM [Na$^+$])
while for single-stranded RNA (poly-U) we considered the recent
results of ref.~\cite{Koen} (solutions of 500, 300, 100, 50, 10 and 5
mM [Na$^+$]). The fits of the experimental data is carried out through
the standard procedure of minimizing the $\chi^2$ over the TC
parameters: the contour length, $L_c$, chain granularity, $a$ and
chain thickness, $\Delta$. For the calculation of $\chi^2$ we have
estimated the effective uncertainty on the force measurements, taking
the relative extension as the independent variable, directly from the
large data sets of poly-U (more than 400 points for each set). For
dsDNA, owing to the more limited number of points (about 20-25) we
propagated the declared experimental uncertainty on extension.

\begin{figure}[h]
\centering
 \includegraphics[width=2.5in]{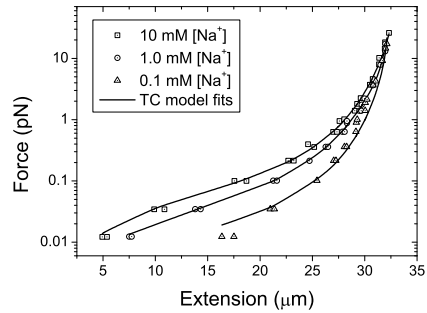}
\caption{Thick-chain fit of the experimental data (squares, circles
and triangles) on dsDNA in solutions of various ionic strengths. The
best-fit parameters are provided in Table
\ref{table:DNA}.}\label{fig:DNA}
\end{figure}

\begin{figure}[h]
    \subfigure[]
    {
        \includegraphics[width=2.5in]{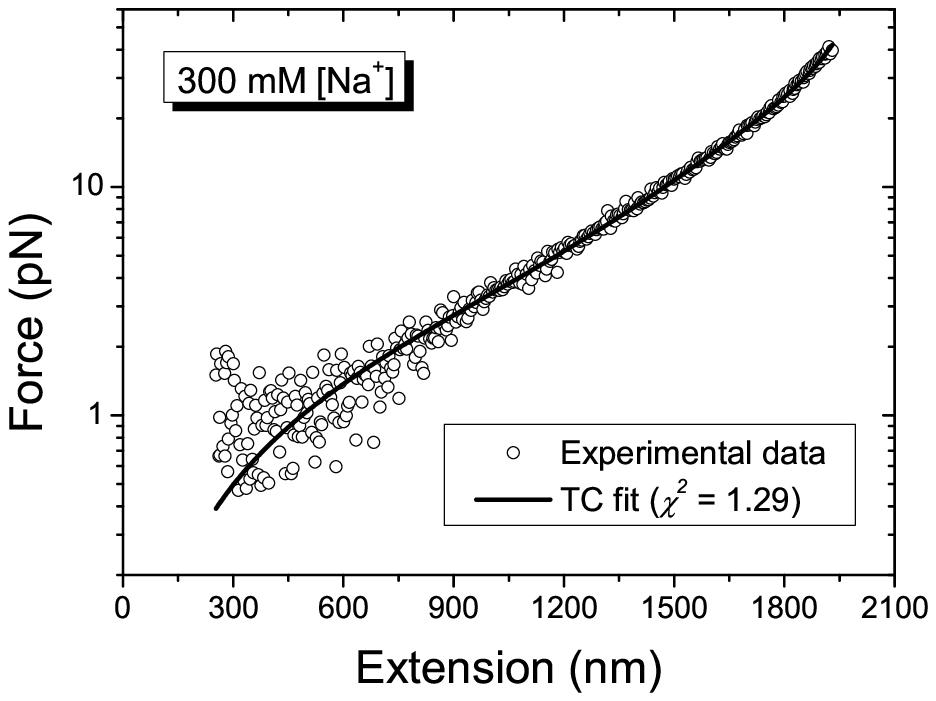}
    }
    \subfigure[]
    {
        \includegraphics[width=2.5in]{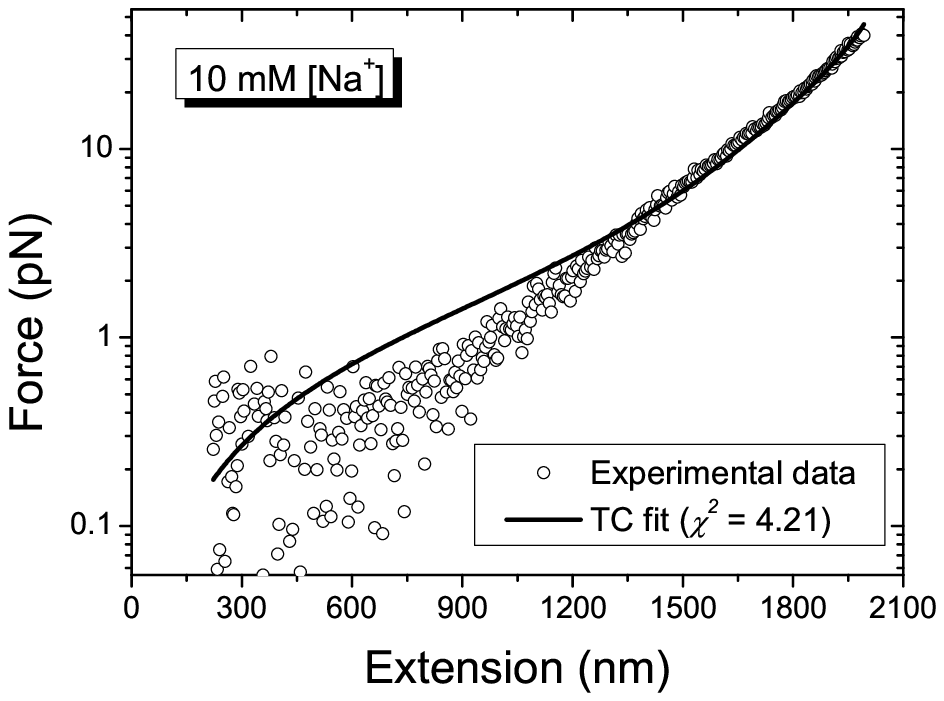}
    }
\caption{Application of the thick-chain model to Poly-U stretching
data for (a) 300 mM [Na$^+$] and (b) 10 mM [Na$^+$]. Experimental data
are shown as open circles, the fit with the TC is denoted with a solid
line. The best-fit parameters for the TC are provided in Table
\ref{table:RNA}.}\label{fig:RNA}
\end{figure}

The results of the fit procedures are provided in graphs~\ref{fig:DNA} and~\ref{fig:RNA} and tables~\ref{table:DNA} and~\ref{table:RNA}.

\begin{table}[h]
 \centering
\begin{tabular}{|c|c|c|c|c|c|}
\hline
 \multicolumn{6}{|c|}{dsDNA}\\
 \hline
   [Na$^+$]  & $a$  &  $\Delta$  &  $\xi_p$   &  $L_c$      &  $\chi^2$ \\
     (mM)  & (nm) &  (nm) &  (nm)      &  (nm)   & \\
 \hline
10     & 9  $\pm$  2  & 12 $\pm$  2 &  55 $\pm$ 2     & 32.9  $\pm$  0.2  &  0.6 \\
1.0    & 25 $\pm$  7  & 26 $\pm$  4 &  94 $\pm$ 4     &  32.3 $\pm$  0.2  &   0.3\\
0.1    & 36 $\pm$  16 & 48 $\pm$  13&  242 $\pm$ 14   &  32.3 $\pm$  0.2  &  1.9\\
 \hline
\end{tabular}
\caption{Best-fit parameters obtained by applying the parametric
force-extension expressions of the TC model to the experimental data
on DNA in solutions of various ionic strengths.}
\label{table:DNA}
\end{table}

\begin{table}[h]
  \centering
\begin{tabular}{|c|c|c|c|c|c|c|}
\hline
  \multicolumn{6}{|c|}{poly-U}\\
  \hline
[Na$^+$] & a  &$\Delta$ &
$\xi_p$ &   $L_c$ & $\chi^2$\\
 (mM)  & (nm)  &(nm) &
(nm)& (nm)& \\
\hline
500     &     1.10    $\pm$   0.01    & 0.64    $\pm$   0.01    &   0.80    $\pm$   0.01    &   2126 $\pm$   4    &   1.49    \\
300     &     1.09    $\pm$   0.02    & 0.65    $\pm$   0.01    &   0.91    $\pm$   0.01    &   2123 $\pm$   5    &   1.29    \\
100     &     1.08    $\pm$   0.03    & 0.68    $\pm$   0.02    &   1.11    $\pm$   0.01    &   2117 $\pm$   8    &   1.05    \\
50      &     1.11    $\pm$   0.03    & 0.73    $\pm$   0.02    &   1.30    $\pm$   0.01    &   2134 $\pm$   8    &   1.33    \\
10      &     1.33    $\pm$   0.08    & 0.94    $\pm$   0.04    &   1.94    $\pm$   0.03    &   2138 $\pm$   12   &   4.21    \\
5       &     1.41    $\pm$   0.10    & 1.03    $\pm$   0.05    &   2.26    $\pm$   0.05    &   2142 $\pm$   12   &   4.34    \\
\hline
\end{tabular}
\caption{Best-fit parameters obtained by applying the parametric
force-extension expressions of the TC model to the experimental data
on poly-U in solutions of various ionic strengths.}\label{table:RNA}
\end{table}

It is particularly appealing that, over the about 400 data points
available for poly-U, the $\chi^2$ associated to the thick chain is
very close to 1 for the set of measurements carried out for [Na$^+$]
in the 50 to 500 mM range. In the case of the two smallest
concentrations, [Na$^+$]= 5 and 10 mM, a significant increase of the
$\chi^2$ is observed. The same is true for the fit of DNA measurements
carried out in 0.1 mM [Na$^+$]. The worsening of the TC performance
consequent to the increase of the screening length is reflected, among
other effects, by the progressive importance of accounting for
overstretching \cite{podg00,williams,Koen}. Accordingly, the fitting
parameters obtained at the lowest salt concentrations can be expected
to change upon the introduction of a stretching modulus.

For both DNA and RNA the viability of the structural and elastic
parameters of the TC fit can be compared against those obtained by
different models and physical approaches. The most natural term of
comparison for the elastic response is provided by the widely-employed
WLC, which is the common reference model for determining the
persistence length and contour length of several types of
biopolymers. It is appealing that both $\xi_p$ and $L_c$ given by the
TC for dsDNA and poly-U in Tables~\ref{table:DNA} and~\ref{table:RNA}
are in close agreement with the same quantities given by the WLC, the
relative difference being typically less than 5\% for both quantities.
For dsDNA, the values of $\xi_p$ as a function of [Na$^+$] are
compatible with the ones found by Baumann {\em et al.}~\cite{WLC4} and
those predicted by Lee and Thirumalai~ \cite{thiru}. The values of
$\xi_p$ for poly-U are, on the other hand, consistent with the WLC
results found by Seol {\em et al.}~in the original analysis of the
data set provided to us~\cite{Koen}. Also, the order of magnitude of
the granularity parameter, $a$, providing the best fit for various
[Na$^+$] concentrations is consistent with the discretization length
of the piece-wise cylindrical model optimised for DNA by Vologodskii and Frank-Kamenetskii ~\cite{volog92}.

It is important to point out that the best fits of the TC model yield
a $\chi^2$ that is about one half the WLC one. This way reflect the
fact that the WLC has one less parameter (the granularity) with
respect to the TC model. However, the difference also stems from the
very different functional dependence of $f(x)$ in the two models. A
noteworthy illustration of this fact is provided by the use of the
Kratky-Porod (KP) model which represents a discretised chain with
bending rigidity (e.g. a WLC endowed with a granularity
parameter). The presence of the additional parameter, which allows the
comparison with TC on a physically-equal footing, does not decrease
appreciably the $\chi^2$ which remains about twice the TC one. A
considerable improvement over the WLC/KP fit is however possible upon
the inclusion of the local thickness effects, that is through the
LTC+BR model of eqns.~\ref{eqn:LTCBR}
and~\ref{eq:TC_KP_simplest}. The model is particularly interesting
because it represents the simplest way of combining, in an approximate
but analytical way, bending rigidity and thickness effects (within the
piece-wise cylindrical model the inclusion of bending rigidity was
recently discussed in ref.~\cite{bensimon_braid}). Despite the absence
of non-local self-avoidance the LTC+BR model has a $\chi^2$
performance that, on average, improves the TC one by 20\%. In fact,
for the three concentrations of [Na$^+$] reported in
Table~\ref{table:DNA} one obtains the following $\chi^2$ values: 0.6,
0.4 and 1.0. The associated thickness values (being approximately 14,
26, and 48 nm, for 10, 1.0 and 0.1 mM [Na$^+$], respectively) are
consistent with the full TC determination, while the bending rigidity
parameters are within 50\% of those of the ``bare'' WLC.  These
results indicate the benefit of supplementing the ordinary bending
rigidity with the self-avoidance originating from the chain
thickness. Owing to the fact that the combined model possesses a
larger parameter space we will postpone to future work the detailed
characterization of the model.

We finally discuss the dependence of the effective thickness,
$\Delta$, on the concentration of counterions in solution (the
apparent diameter of the polymer is simply given by $2 \Delta$). For
both DNA and poly-U, the effective radius shows a monotonic decrease
for increasing concentrations and indicates reaching a saturated value
(different for each type of the polymers), namely the inferred
``bare'' thickness. In particular, over the range of 0.1 to 10 mM
[Na$^+$] the apparent radius of dsDNA decreases from 48 nm to 12 nm.
For ssRNA the decrease is instead, 0.96 nm to 0.63 nm over the range
of 5 to 500 mM [Na$^+$]. As we anticipated in the introduction, the
theoretical results derived by Stigter and Odijk~\cite{stigter77,odijk90}
have been previously used to predict the effective diameter of
double-stranded DNA.  The available theoretical predictions based on
the approach of refs.~\cite{stigter77,stigter93,odijk90} yield dsDNA {\em radii} of
8.1 nm for 10 mM [Na$^+$], 27.1 nm for 1 mM [Na$^+$] and 96.3 for 0.1
mM [Na$^+$]. Given the very diverse nature of the approach and
approximations of the present study and of refs.~\cite{stigter77,odijk90},
it is pleasing that the two determinations of the effective thickness
are in reasonable agreement.

\begin{figure}[h]
\begin{center}
\includegraphics[width=3.5in]{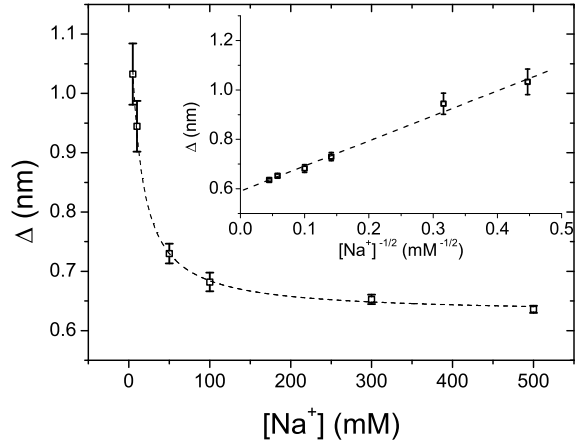}
\end{center}
\caption{Effective radius, $\Delta$, of poly-U as a function of
ionic strengths. The dashed curve is a visual guideline obtained
from a spline interpolation of the points obtained from the TC fits.
The approximate linear dependence of $\Delta$ on
$1\over\sqrt{[Na^+]}$ (with [Na$^+$] expressed in molar units) is
shown in the inset.}\label{fig:Deltaion}
\end{figure}

We finally turn to the case of poly-U. Owing to its high flexibility,
even short stretches of RNA cannot be modeled as stiff rods, and hence
the theoretical analysis of ref.~\cite{stigter77} cannot be used to
generate a comparison term for the effective radius, $\Delta$, of
poly-U. It was pointed out by Stigter that for DNA the electrostatic
contribution to the effective {\em radius}, though exceeding by
several times the bare one, was of the same order of the Debye
screening length, $\lambda_D$ (the ``proportionality factor'' ranging
from about 4 to 8) \cite{stigter77}.  Also in the case of RNA, the
Debye length, which is proportional to $1/\sqrt{[{\rm Na^+}]}$ is a
useful concept for rationalising the dependence of the effective
radius on the ionic strength. In fact, within the explored range of
[Na$^+$], $\Delta$ appears to increase linearly with $1/\sqrt{[{\rm
Na^+}]}$, as visible in the inset of Fig. \ref{fig:Deltaion}
(b). Assuming the validity of such linear relationship, the ``bare''
radius of poly-U is estimated by extrapolating $\Delta$ for vanishing
$\lambda_D$.  One obtains $\Delta_{bare} \approx 0.59 \pm 0.01 nm$,
which compares well with the nominal value of about 0.5 nm observed in
crystallographic structures of (non-hydrated) poly-U fragments (from
PDB structure 1I5L). Incidentally we mention that also the value of
$a$ agrees with the nominal size of the modular poly-U units which is
about 0.6 nm. Several efforts have been devoted in the past to
establishing how the apparent persistence length of polyelectrolytes
is affected by the interplay between the bare persistence length, the
linear charge density and the ionic strength
\cite{Ullner97,Reed91,thiru}. Different manners of functional
dependence of $\xi_p$ on $\lambda_D$ are, in fact, observed for
polyelectrolytes with different degree of flexibility, such as
double-stranded and single-stranded DNA. It is, therefore, interesting
to compare the dependence of the effective diameter of DNA and
poly-U. The most notable difference is that the proportionality factor
between the electrostatic contribution to the effective radius and the
Debye length is of the order of 0.1, therefore appreciably smaller
than for the case of dsDNA.

\section*{Conclusions}

Extensive stochastic sampling techniques were used to characterize
the behaviour of an inextensible thick polymer subject to a
stretching force. The extension versus force response found
numerically was parametrised in terms of the polymer structural
parameters (thickness and monomer length) and captured by an
analytic expression. The resulting formula was used to fit
experimental data obtained from stretching measurements of DNA and
poly-U in solutions of different ionic strength. This represents a
novel and physically-appealing route to extract the apparent
structural parameters of polyelectrolytes starting merely from their
elastic response. The inferred effective diameter for DNA was found
to be in satisfactory agreement with the estimates obtained by
Stigter through an unrelated approach. In particular, the
electrostatic contribution to the effective DNA radius was found to
be several times larger than the Debye screening length,
$\lambda_D$. Also for the much-more flexible poly-U chain it is
observed that the effective radius strongly depends on the ionic
strength, having an approximately linear dependence on $\lambda_D$
within the available ranges of [Na$^+$]. At variance with dsDNA,
however, the electrostatic contribution to the effective radius was
found to be an order of magnitude smaller than the Debye screening
length.

\vskip 1.0cm

{\it Acknowledgements:} We are indebted to K. Visscher for having
provided us with the experimental data on poly-U and for useful
comments. We thank T. Odijk for valuable suggestions and are grateful
to D. Marenduzzo for several discussions and the careful reading of
the manuscript.

\end{document}